\documentclass[showpacs,aps,prb,amsmath,amssymb,superscriptaddress]{revtex4}
\usepackage{graphicx}
\usepackage{dcolumn}
\usepackage{bm}
\begin{document}

\baselineskip=0.8cm
\renewcommand{\thefigure}{\arabic{figure}}
\title{Comparative study of screened inter-layer interactions
in the Coulomb drag effect in bilayer electron systems}
\author{R. Asgari}
\affiliation{Institute for Studies in Theoretical Physics and
Mathematics, Tehran 19395-5531, Iran}
\author{B. Tanatar}
\affiliation{Department of Physics, Bilkent University,
Bilkent, 06800 Ankara, Turkey}
\author{B. Davoudi}
\affiliation{D{\'e}partment de Physique and Centre de Recherche en
Physique du Solide, Universit{\'e} de Sherbrooke, Sherbrooke,
Qu{\'e}bec, Canada J1K 2R1}

\begin{abstract}
Coulomb drag experiments in which the inter-layer resistivity is
measured are important as they provide information on the Coulomb
interactions in bilayer systems. When the layer densities are low
correlation effects become significant to account for the
quantitative description of experimental results. We investigate
systematically various models of effective inter-layer interactions
in a bilayer system and compare our results with recent experiments.
In the low density regime, the correlation effects are included via
the intra- and inter-layer local-field corrections. We employ
several theoretical approaches to construct static local-field
corrections. Our comparative study demonstrates the importance of
including the correlation effects accurately in the calculation of
drag resistivity. Recent experiments performed at low layer
densities are adequately described by effective inter-layer
interactions incorporating static correlations.
\end{abstract}

\maketitle

\section{Introduction}

In the last decade transport properties of dilute two-dimensional
(2D) electron and hole systems have amassed a great interest. Much
of the excitement and controversy is centered around the temperature
dependence of resistivity which appears to exhibit metallic behavior
at high densities and insulating behavior at low
densities.\cite{kravchenko} In bilayer systems in which the barrier
separating the coupled quantum wells is large enough so that
tunneling effects are negligible, the inter-layer resistivity has
been measured for more than a decade.\cite{gramila} In this
so-called drag effect the momentum transfer between the layers is
measured.\cite{rojo}
In contrast to the single layer resistivity
which shows a nontrivial interplay between interaction and disorder
effects near the metal-insulator transition\cite{punnoose},
the inter-layer resistivity is largely determined by the long range
Coulomb scattering (as long as the single layer densities are
away from metal-insulator transition region).
Therefore Coulomb drag experiments provide valuable information on
the intra- and inter-layer electron-electron interactions especially
when the layer densities are lowered.

Over the years there has been a number of Coulomb drag experiments
at zero magnetic field using different samples and probing different
parameter regimes. The main parameters entering a drag experiment
set-up are the layer density $n$ which may be related to the
dimensionless coupling strength $r_s$ (for the definition of $r_s$
see Section II), the separation distance between the layers $d$ and
the Fermi temperature $T_F$. Hill {\it et al}.\cite{hill} measured
drag resistivity $\rho_D$ in an electron bilayer system at densities
corresponding to $1.13\lesssim r_s\lesssim 1.57$ and high
temperatures $T\sim T_F$. The observed peak in $\rho_D$ around
$T\approx T_F/2$ was attributed to the contribution of plasmons. In
fact, the experimental results were regarded as an indirect evidence
for the existence of acoustic and optical plasmons in a bilayer
system.\cite{flensberg-hu} Similar experiments were also performed
by Noh {\it et al}.\cite{nohplas} confirming plasmon effects on the
drag resistivity and revealing the importance of possible dynamic
correlations even though the layer densities were such that
$r_s\approx 1.48$ where the strong coupling effects are not
expected. More recent experiments by Kellogg {\it et
al}.\cite{kellogg_02} used samples with layer densities reaching
$r_s\approx 4.3$ and $k_Fd\sim 1$ where $d$ is the center-to-center
well separation. In contrast to the above experiments, Pillarisetty
{\it et al}.\cite{pillarisetty_02} measured frictional drag between
two dilute 2D hole layers in which the $r_s$ values were in the
range $19\le r_s\le 39$.

On the theoretical side, the drag resistivity has first been
formulated within the random-phase approximation (RPA) for the layer
density-response functions and inter-layer effective
interaction.\cite{jauho,zheng_93} Here and most subsequent works
treat the inter-layer effective interaction as given by the bare
inter-layer Coulomb interaction screened by the bilayer system
dielectric function. Importance of dynamical correlations is noticed
even at the RPA level since the difference between the static and
dynamic screening function brings quantitative changes to the drag
resistivity.\cite{jauho} At larger $r_s$ values when the correlation
effects become significant one should go beyond the RPA. One way to
do this in a physically motivated way is through the local-field
corrections to the RPA form of the screening function. The simplest
form of the local-field corrections is the Hubbard approximation
which was used by Hill {\it et al}.\cite{hill} to analyze their
data. A much widely used local-field corrections are calculated
within the self-consistent field approximation scheme of Singwi {\it
et al}.\cite{singwi_68} (STLS). They have been incorporated in the
evaluation of the drag resistivity by {\'S}wierkowski {\it et
al}.\cite{swierkowski_97}. In connection with the Kellogg {\it et
al}. experiments\cite{kellogg_02}, Yurtsever {\it et
al}.\cite{yurtsever_03} pointed out that STLS local-field
corrections yield a poor representation and suggested the use of a
different effective interaction originally developed by Kukkonen and
Overhauser\cite{ko_79} and Vignale and Singwi.\cite{vignale_85}
Recently, Badalyan {\it et al}.\cite{badalyan} employed frequency
dependent local-field corrections in the long-wavelength limit
($q\rightarrow 0$) obtained from dynamical exchange-correlation
kernel in the context of density functional theory.

In this work we investigate systematically the effect of the form of
screened inter-layer interaction on the temperature dependence of
drag resistivity. We calculate the drag resistivity employing
several models for the inter-layer interaction and compare their
behavior with the experimental results of Kellogg {\it et
al}.\cite{kellogg_02} which provide a useful test at low density. As
input to various theoretical models of inter-layer interaction we
consider several constructs of local-field corrections. Our
calculations reveal the importance of the choice of inter-layer
interaction model and the significant role played by the local-field
corrections.

The rest of this paper is organized as follows. In Sec.\,II, we
introduce the models for inter-layer interaction that enters the
drag resistivity. We then outline the calculation of local-field
corrections in various approaches. Section III contains our
numerical calculations of drag resistivity and comparison of models
with experimental data. We conclude in Sec.\,IV with a brief
summary.

\section{Theoretical Approach}
\label{sect:bh}

We consider a double-quantum-well structure with $d$ as the
center-to-center well separation such that there is no tunneling
between them and $L$ as the width of the quantum wells. Each layer
is characterized by the dimensionless coupling constant $r_s
a^*_B=1/\sqrt{\pi n}$ where $n$ is the areal density,
$a^*_B=\hbar^2\epsilon/(m^* e^2)$ is the effective Bohr radius,
$\epsilon$ and $m^*$ being the background dielectric constant and
electron band effective mass. Each layer has only one type of charge
carrier, i.e. electrons, although our theoretical formulation could
be applicable to hole-hole and electron-hole layers with suitable
changes. In the case of electron-hole bilayers the prospect of
formation of an excitonic state\cite{eh_senatore} and its detection through
drag experiments\cite{eh_prospect} requires a new formulation of the
effective inter-layer interaction which we do not address here.
However, correlations in electron-hole bilayers and their effects
on drag resistivity can be studied using the improved inter-layer
models we shall describe below.
The motion of the carriers is free along the $xy$ plane and
under the action of a double-well potential profile in the
$z$-direction only the lowest subband in each quantum well is
occupied. For this aim, temperature should be less than the
difference between excited energy level and the ground state energy
in quantum well. This yields $T < 3(r_s a^*_B/L)^2 T_F/16$.
Furthermore, the bilayer system is assumed to be embedded in a
uniform neutralizing positive background charge. The unscreened
Coulomb interaction potential, in Fourier space, between the
electrons in $k$th and $l$th layers is given by $v_{kl}(q)=v_q
F_{kl}(qL)$. Here, $v_q=2\pi e^2/(\epsilon q)$ and $F_{kl}$ are
infinite quantum-well form factors taking the finite width effects
into account which are given by ~\cite{jauho}
\begin{eqnarray}\label{form}
F_{kk}(x)&=&\frac{3x+8\pi^2/x}{x^2+4\pi^2}-
\frac{32\pi^4[1-\exp(-x)]}{x^2(x^2+4\pi^2)^2}\nonumber\\
F_{kl}(x)&=&\frac{64\pi^4\sinh^2(x/2)}{x^2(x^2+4\pi^2)^2}\exp(-qd)\, .
\end{eqnarray}
We note that most theoretical
calculations\cite{jauho,flensberg-hu,swierkowski_97} adopt the
infinite quantum-well model to account for the width effects,
whereas a better way would be to calculate the Coulomb matrix
elements using envelope functions $\phi_n(z)$ determined
self-consistently from the Poisson and Schr{\"o}dinger
equations.\cite{kainth}

The drag resistivity (or as it is also
called transresistivity) $\rho_D$ of an electron system at
temperature $T$ has been obtained in a variety of theoretical
models. These include diagrammatic perturbation
theory~\cite{jauho,kamenev_95}, the Boltzmann
equation~\cite{flensberg_95} and the memory function
formalism~\cite{zheng_93,swierkowski_97}. In a drag experiment one
applies an electric field $E_1$ to layer 1 (drive layer) creating a
current to flow with current density $J_1$. This sets up an electric
field $E_2$ in layer 2 (drag layer) where no current is allowed to
flow. The drag resistivity is defined as $\rho_D=E_2/J_1$ and the
microscopic calculations relate this quantity to the rate of change of
momentum between the layers, as
electron-electron inter-layer interactions transfer
momentum from the drive layer with carrier density $n_{1}$ to the
drag layer with density $n_{2}$.

Theoretical considerations lead
to the same expression for $\rho_D$ in terms of the effective
inter-layer interaction and the density-response function of the
single layers. When the effective inter-layer
interaction treated perturbatively, $\rho_D$ is given as
\begin{equation}\label{drag}
\rho_D= -\frac{\hbar^2}{8 \pi^2 e^2 n_1 n_2 k_B T}\int_0^\infty q^3
dq \int_0^\infty d\omega \frac{\left|W_{12}(q,\omega)\right|^2~ \Im
m\chi^0_1 (q,\omega,T) ~\Im m\chi^0_2 (q,\omega,
T)}{\sinh^2(\hbar\omega/2k_B T)}~,
\end{equation}
where $\chi^0_i(q,\omega)$, ($i=1$ or 2) is the non-interacting
linear response corresponding to the drive and drag layer which
shows the charge density fluctuations in a given layer at finite
temperature and $W_{12}(q,\omega)$ is the effective inter-layer
interaction.

An important ingredient which is needed to calculate $\rho_D$ is the
electron-electron inter-layer interaction, $W_{12}(q,\omega)$. The
effective electron-electron interaction for a two-component system
given by a $2\times 2$ matrices and in random-phase approximation
(RPA) it is given by
\begin{equation}
\hat{W}_{RPA}(q,\omega)=\hat{v}(q)+
\hat{v}(q)\hat{\chi}(q,\omega)\hat{v}(q)~,
\end{equation}
where $\hat{\chi}(q,\omega)$ defined in terms of the non-interacting
charge-charge response function and Coulomb interactions.

To take into account the effect of correlations more clearly, which
are more important in the strongly correlated regime where $r_s$
becomes large, we need more sophisticated approaches. For this
purpose, we introduce here other approximation scheme for
$W_{12}(q,\omega)$ proposed by {\'S}wierkowski {\it et
al}.~\cite{swierkowski_97,badalyan} (SSG) where
\begin{equation}\label{wssg}
\hat{W}_{SSG}(q,\omega)=\hat{v_{eff}}(q)+
\hat{v}_{eff}(q)\hat{\chi}(q,\omega)\hat{v}_{eff}(q)~,
\end{equation}
where $v^{ij}_{eff}(q)=v_{ij}(q)(1-G_{ij}(q))$ are the effective
Coulomb interactions and $G_{ij}(q)$ are intra- and inter
local-field corrections (LFC) which take into account multiple
scattering to infinite order between all components of the plasma
compared with the RPA where these effects are neglected.

A more detailed analysis, which accounts for the vertex corrections
associated with charge-charge fluctuation, was carried out for an
electron gas (EG) in Refs.\,\onlinecite{vignale_85,richardson_97,
yurtsever_03}, where Kukkonen-Overhauser-like effective inter-layer
interaction potential~\cite{ko_79} were obtained by different
approaches. In this scheme we have
\begin{equation}\label{wvs}
\hat{W}_{VS}(q,\omega)=\hat{v}_{eff}(q)+
\hat{v}_{eff}(q)\hat{\chi}(q,\omega)\hat{v}_{eff}(q)-\hat{U}~,
\end{equation}
with the elements of $\hat{U}$ defined by $v_{ij}(q)G_{ij}(q)$.
The form of $W_{12}(q,\omega)$ within the Vignale and Singwi (VS)
approach is similar to that in the self-consistent field approach of
Singwi {\it et al}.~\cite{swierkowski_97,singwi_68} (SSG) except for
the last term. More clearly, the inter-layer interaction in
Eq.\,(\ref{wvs}) is given by~\cite{vignale_85,richardson_97}
\begin{equation}\label{w12}
W_{12}(q,\omega)|_{VS}=
\frac{v_{12}(q)(1-G_{12}(q))}{\Delta(q,\omega)}-
v_{12}(q)G_{12}(q)\, ,
\end{equation}
where
\begin{equation}
\Delta(q,\omega)=[1-v_{11}(q)(1-G_{11}(q))\chi^0_1(q,\omega, T))
(1-v_{22}(q)(1-G_{22}(q))\chi^0_2(q,\omega,
T)]-[v_{12}(q)(1-G_{12}(q)]^2 \chi^0_1(q,\omega,
T)~\chi^0_2(q,\omega, T)\, .
\end{equation}
Here $\chi^0_k(q,\omega,T)$ is non-interacting charge-charge response
function at finite temperature.\cite{flensberg-hu}

Another approximation scheme for screened bilayer 2D
electron-electron interaction is proposed by Zheng and
MacDonald~\cite{zheng}(ZM). In this scheme the screened
electron-electron interaction potential is given as
\begin{equation}\label{wzm}
\hat{W}_{ZM}(q,\omega)=\left[ 1-\hat{\chi}^0(q,\omega,
T)\hat{v}_{eff}(q)\right]^{-1}\hat{v}(q)\, .
\end{equation}
This is derived essentially from a two-component generalization of
the vertex function that enters in self-energy in the so-called
$GW\Gamma$ approximation. However, because of the matrix nature of
two-component systems there seems to be some ambiguity in such a
construction. Note, for instance, that $\hat{W}_{ZM}$ is not a
symmetric matrix for unmatched bilayer systems. Finally, we remark
that VS, SSG and ZM forms of the effective electron-electron
interactions reduce to RPA if the LFCs are omitted.

As it is clear from Eqs.\,(\ref{wssg}), (\ref{wvs}) and (\ref{wzm})
the local-field corrections are the fundamental quantities for an
evaluation of the effective electron-electron interaction. Here, we
intend to examine the inter-layer potential of the Coulomb bilayer
system by including correlation effects. To this purpose, we made
use of the STLS approach both at zero (STLS0) and finite temperature
(STLS) schemes. The STLS theory embodies correlations beyond the RPA
approach and as an important improvement. In this approach the
static LFC that accounts for correlation effects among carriers in
the layers $k$ and $l$ are given by:
\begin{equation}\label{LFC}
G_{kl}(q)=-\frac{1}{n}\int \frac{d{\bf q'}}{(2\pi)^2}~\frac{{\bf
q.k}}{q^2}\frac{v_{kl}(q')}{v_{kl}(q)}~[S_{kl}(|{\bf
q-q'}|)-\delta_{kl}]~,
\end{equation}
where $S_{kl}(q)$ is a static structure factor. The equations of
motion for the Wigner distribution functions in a bilayer coupled
with the linear-response theory yield in the Singwi {\it et
al}\cite{singwi_68} approach the following expression for the
density-density response functions:
\begin{equation}\label{chi}
\chi_{kl}(q,\omega)=\frac{\chi^0_k(q,\omega,
T)\left\{\delta_{kl}+
(-1)^{\delta_{kl}}v_{kl}(q)(1-G_{kl}(q))\chi^0_l(q,\omega,
T)\right\}}{\Delta(q,\omega)}~.
\end{equation}
The fluctuation-dissipation theorem leads to the static structure
factor for a bilayer at finite temperature
\begin{equation}\label{Sq}
S_{kl}(q)=-\frac{\hbar}{\pi\sqrt{n_k~n_l}}\int d\omega \Im
m\chi_{kl}(q,\omega)~\coth\left(\frac{\hbar\omega}{2k_BT}\right)~.
\end{equation}
Equations (\ref{LFC}), (\ref{chi}) and (\ref{Sq}) are solved
numerically in a self-consistent way for $G_{kl}(q)$ both at zero
and finite temperature cases separately.

Another sophisticated method is based on Fermi hypernetted-chain
approach (FHNC). Our strategy follows a similar approach to our
recent works, Ref.\,[\onlinecite{asgari_2}] which uses accurate intra- and
inter-layer static structure factors to build the local-field corrections.
For this purpose we implement the self consistent Fermi
hypernetted-chain approach~\cite{lantto,zab,report} at zero
temperature in order to calculate the intra- and inter-layer
static structure factors incorporating the finite thickness effects in a
quantum well. The latter effects are known to be important for the
adequate description of the drag resistivity from a number of
calculations.\cite{jauho,swierkowski_97,flensberg-hu,badalyan}
In what follows we explain the FHNC approximation and
then outline our method to obtain the static local-field
corrections, $G_{ij}(q)$, at zero temperature.

With the zero of energy taken at the chemical potential, the
formally exact differential equation for the pair-correlation
function\cite{saeed}, $g_{\alpha\beta}(r)$, and following
Chakraborty\cite{tc} using the two-component plasma Jastrow-Slater
variational theory involving FHNC approximation, reads
\begin{equation}\label{EL}
\left[-\frac{\hbar^2}{m}\nabla^2+V^{eff}_{\alpha\beta}(r)\right]
\sqrt{g_{\alpha\beta}(r)}=0\quad,
\end{equation}
where $m$ is electron mass and
$V^{eff}_{\alpha\beta}(r)=v_{\alpha\beta}(r)+W_{\alpha\beta}^B(r)+
W_{\alpha\beta}^F(r)$. In Eq.\,(\ref{EL}) we decompose the effective
potential into three terms $v_{\alpha\beta}(r)$, $W_{\alpha\beta}^B$
and $W_{\alpha\beta}^F$ of which the last two terms take into
account correlation and exchange effects respectively, we substitute
to the direct boson potential $W_{\alpha\beta}^B$ the one calculated
by Chakraborty \cite{tc} for a two component Bose system using the
static structure factors $S_{\alpha\beta}(k)$ of a Fermi system:
\begin{equation}\label{correlation}
\left\{
\begin{array}{l}
W^B_{\alpha\alpha}(k)={\displaystyle -
\frac{\hbar^2k^2}{4m n_{\alpha}}\left[2S_{\alpha\alpha}(k)-3+
(S^2_{{\bar\alpha}{\bar\alpha}}(k)+
S^2_{\alpha{\bar \alpha}}(k))/\Gamma^2(k)\right]}\\
\\
W^B_{\alpha{\bar \alpha}}(k)={\displaystyle
-\frac{\hbar^2k^2}{4m\sqrt{n_{\alpha}n_{\bar \alpha}}}
\left[2S_{\alpha{\bar \alpha}}(k)-S_{\alpha{\bar
\alpha}}(k)(S_{{\bar \alpha}{\bar
\alpha}}(k)+S_{\alpha\alpha}(k))/\Gamma^2(k)\right]}
\end{array}
\right.
\end{equation}
Here $S_{\alpha\beta}(k)=\delta_{\alpha\beta}+
\sqrt{n_{\alpha}n_{\beta}}\int[g_{\alpha\beta}(r)-1]
\exp(i{\bf k}\cdot{\bf r})d{\bf r}$ is the static structure factor
and
\begin{equation}\label{Delta}
\Gamma(k)=S_{11}(k)S_{22}(k)-S^2_{12}(k)\, .
\end{equation}
Turning to the exchange term $W^F_{\alpha\beta}$, it is itself
defined in order to make Eq.\,(\ref{EL}) exact and has a very
complicated expression within the FHNC.\cite{lantto,zab,report}
However, in dealing with a one-component electron fluid, Kallio and
Piilo\cite{kp} have proposed a simple and efficient way to account
for the antisymmetry of the fermion wave function. Their argument is
immediately generalized to our two-component Fermi fluid, and leads
to the requirement that, in the high density regime in both layers,
the Hartree-Fock pair distribution functions are solution of
Eq.\,(\ref{EL}). Following this prescription, it turns out that
$W^F_{\alpha\beta}(k)$ is given by,
\begin{equation}\label{exchange}
W^F_{\alpha\alpha}(k)=\int\frac{\hbar^2}{m}~\frac{\nabla_{\bf r}^2
\sqrt{g^{\rm \scriptscriptstyle HF}_{\alpha\alpha}(r)}}{\sqrt{g^{\rm
\scriptscriptstyle HF}_{\alpha\alpha}(r)}}e^{i{\bf kr}}d{\bf r}
+\frac{\hbar^2k^2}{4mn_{\alpha}}\left[2S^{\scriptscriptstyle\rm
HF}_{\alpha\alpha}(k)-3+\left(\frac{1}{S^{\scriptscriptstyle\rm
HF}_{\alpha\alpha}(k)}\right)^2\right]~,
\end{equation}
and $W^F_{\alpha{\bar\alpha}}(k)=0$. In Eq.\,(\ref{exchange}) we
used the fact that the Coulomb term in Eq.\,(\ref{EL}) becomes
negligible in the Hartree-Fock limit with respect to the kinetic
term.

Although the expression for the exchange potential in
Eq.\,(\ref{exchange}) is correct only for weakly coupled Fermi
fluids, we shall assume in the following that it can yield useful
results in our self-consistent calculations of the pair distribution
functions with increasing coupling strength~\cite{saeed}. As a broad
qualitative argument in support of this assumption we may remark
that the role of the statistics is expected to weaken with
increasing coupling strength making the correlation term dominate on
the exchange
one. In Eq.\,(\ref{exchange}) $S^{\rm\scriptscriptstyle
HF}_{\alpha\alpha}(k)$ and
$g^{\rm \scriptscriptstyle HF}_{\alpha\alpha}(r)$
are, respectively, the static structure factor
and the intra-layer pair distribution functions in the Hartree-Fock
approximation (HF), namely
\begin{equation}\label{shf}
S^{\rm \scriptscriptstyle HF}_{\alpha\alpha}(k)=
\frac{2}{\pi}\left[\arcsin\left(\frac{k}{2k_{F_{\alpha}}}\right)
+\frac{k}{2k_{F_{\alpha}}}\sqrt{1-
\left(\frac{k}{2k_{F_{\alpha}}}\right)^2}\,\,
\right]\vartheta(2k_{F_{\alpha}}-k)
+\vartheta(k-2k_{F_{\alpha}})~,
\end{equation}
and $g^{\rm \scriptscriptstyle HF}_{\alpha\alpha}(r)
=1-2(j_1(rk_{F_{\alpha}})/rk_{F_{\alpha}})^2$
and $g^{\rm \scriptscriptstyle HF}_{\alpha{\bar\alpha}}(r)=1$, where
$j_1$ is a spherical Bessel function, and
$k_{F_{\alpha}}=(2\pi n_{\alpha})^{1/2}$.

It is evident that the insertion of
Eqs.\,(\ref{correlation}-\ref{shf}) into Eq.\,(\ref{EL}) allows a
self-consistent calculation of the pair distribution functions and
of the effective interactions. The fluctuation-dissipation theorem
which is of paramount importance for systems in equilibrium relates
the dynamic susceptibilities defined above to the static structure
factors allows one to determine the local-field corrections once the
static structure factors are calculated by FHNC
approach~\cite{asgari_2}.

\section{Numerical results}

In this section we present our calculations for drag resistivity
$\rho_D$ using the theoretical models described above and compare
them with the recent experimental measurements. We proceed to illustrate
our main numerical results, which are summarized in Figs.~1-7.

The effective inter-layer interaction models which go beyond the RPA
use local-field corrections as input. In Fig.~1, we display the
typical behavior of intra- and inter-layer LFCs $G_{11}(q)$ and
$G_{12}(q)$, respectively. We note that whereas the LFCs in the STLS
approach have a monotone $q$ dependence, FHNC approach yields a
peaked structure. Such a structure in static LFCs is well known from
quantum Monte Carlo simulations\cite{qmc} and it is thought to
represent the correlation effects correctly. Thus, differences in
LFCs will evidently play an important role in the drag resistivity.
We also remark that there is considerable difference between zero
and finite temperature (at $T=1$\,K) STLS calculations especially
for the inter-layer LFC, $G_{12}(q)$. We believe that within our
calculational scheme the FHNC approach yields the most accurate
LFCs. To illustrate our point, we compare the intra- and inter-layer
pair-correlation functions $g_{\alpha\beta}(r)$ resulting from FHNC
calculations and QMC simulations\cite{badalyan} in Fig.\,2(a). We
note that the agreement is very good. The STLS scheme does not
reproduce well the peak structure in $g_{11}(r)$ at this density
which corresponds to $r_s=7.07$. We have also looked at the
inter-layer distance $d$ dependence of the LFCs within the FHNC
approach. Figure 2(b) shows intra- and inter-layer LFCs for various
values of $d$ at a layer density $n=3.1\times 10^{-10}$\,cm$^{-2}$.
We have also used the finite quantum-well widths corresponding to
Kellogg {\it et al}.\cite{kellogg_02} experimental sample. As
expected, the intra-layer LFC $G_{11}(q)$ is not affected much as
$d$ changes, whereas the inter-layer LFC $G_{12}(q)$ becomes smaller
in magnitude as $d$ increases, reflecting the weakened Coulomb
correlations. Similar qualitative results have also been found in a
bilayer STLS calculation.\cite{lerwen}

In Figs.\,3 and 4 we show the calculated drag resistivity as a function of
temperature for various theoretical models of effective inter-layer
interaction (i.e. models denoted as VS, SSG and ZM) with different
LFCs (i.e. schemes denoted as FHNC, STLS and STLS0) at layer
densities $3.1\times 10^{10}$\,cm$^{-2}$ and $3.8\times 10^{10}$\,cm$^{-2}$
and compare them with the
experimental results of Kellogg {\it et al}.\cite{kellogg_02} The
experimental data were obtained for bilayer GaAs-AlGaAs
heterostructures for two identical infinite layers of electrons
separated by $d=$280\,\AA\ and with a double quantum well of widths
$L=180$\,\AA. In all our results, the drag resistivity calculated
within the VS inter-layer potential is larger than the one calculated
within the SSG approximation. It means that the value of $U$ increases
with increasing $G_{12}(q)$, and VS potential in Eq.\,(\ref{wvs})
becomes highly different from the SSG potential given by
Eq.\,(\ref{wssg}). The static LFCs which are constructed within the FHNC
approach together with the electron-electron inter-layer potential
calculated within VS and SSG approaches give results in quite good
agreement with experimental measurements especially in the low
temperature regime below the plasmon-mediated drag. In these figures,
the RPA results underestimate the experimental results. Therefore,
after the inclusion of many-body effects correctly (such as using
FHNC), the drag resistivity is in good quantitative agreement with
experimental results. The LFCs in STLS scheme yield an overestimate
of drag resistivity when it is calculated using the VS and SSG
inter-layer potentials, $W_{12}(q,\omega)$. From the physical point
of view, correlation effects suppress the energies of both the
acoustic and optical plasmons, while finite temperature effects tend
to increase the energies. From this cancelation, the STLS0/SSG gives
results close to the experimental data in comparison to STLS/SSG.
Furthermore, the value of intra-layer LFC at finite temperature at a given
$q<2 k_F$ value, is smaller than the intra-layer LFC at zero temperature
in STLS0 scheme and this yields to have larger plasmon contribution
in drag resistivity when one employs the zero temperature LFCs.
Furthermore, the inter-layer LFC at zero temperature is larger than the
one at finite temperature, thus the drag resistivity in STLS0/VS is
further overestimated than in STLS/VS approach.

Figure 5 shows the log-log plot of the drag resistivity $\rho_D$ as
a function of layer density at $T=1$ and 4\,K. For comparison with
recent calculations of drag resistivity by Badalyan {\it et
al.}\cite{badalyan}, we extract their results from Fig.\,15 (denoted
in the figure by BKVS) and compare them with the results of our
calculations, mainly FHNC/VS and FHNC/SSG. Evidently, our FHNC
calculation produces better agreement with experiment in the whole
range of density compared to all other approximations.

The low temperature behavior of drag resistivity $\rho_D$ for the
samples of Kellogg {\it et al}.\cite{kellogg_02} is important in
understanding the interaction effects.
The low density and close inter-layer spacing such that $k_Fd\lesssim 1$
implies significant contributions of backward scattering processes
to $\rho_D$ and predicts deviations from the usual $T^2$
dependence.\cite{gramila} These deviations expected to be in the form
$\sim T^2\ln{T}$ are difficult to be assessed, but the sensitivity
of $\rho_D$ to layer densities has been noted.
In Fig.\,6 we show the scaled drag resistivity $\rho_D/T^2$ as a function
of temperature for $n=3.8\times 10^{10}$\,cm$^{-2}$ and
$n=2.3\times 10^{10}$\,cm$^{-2}$. The drag resistivity including
the FHNC local-field corrections through the
various screened inter-layer interaction models is compared
with RPA. We note that VS and SSG screened inter-layer interaction
models reproduce the upturn behavior of $\rho_D/T^2$ at low temperature
observed in the Kellogg {\it et al}.\cite{kellogg_02} experiments.
On the other hand, ZM model predicts an opposite behavior in contradiction
with experiments. The increase in $\rho_D/T^2$ at low temperatures
due to exchange-correlation effects were first analyzed by Badalyan
{\it et al}.\cite{badalyan} where they used the static local-field
corrections in this temperature regime.
Similar enhancement in scaled drag resistivity was also obtained
by Hwang {\it et al}.\cite{hwang} in their calculation on bilayer
hole systems in connection with the experiments of Pillarisetty
{\it et al}.\cite{pillarisetty_02} Our comparative study thus
provides information as to which form of screened inter-layer
interaction is more suitable in describing drag experiments at low
density.

Finally, we display the inter-layer distance dependence of the
many-body correlation effects on drag resistivity in Fig.\,7.
When the layer separation is decreased, inter-layer Coulomb
interaction enhances and this leads to an increase in drag
resistivity. Because the Kellogg {\it et al}.\cite{kellogg_02}
experiment did not measure $\rho_D$ for samples with different
$d$ values, we are not able to make a systematic comparison.

In the examples above we have seen that the local-field factors
play a significant role in the quantitative description of the
measured drag resistivity.
It is important to remark that the drag
resistivity is calculated for electron-electron interaction
only and we ignore other scattering processes (i.e. impurities, phonons,
etc.) which may be effective in realistic situations. In general,
the theoretical prediction by the results of
Eq.\,(\ref{drag}) should yield values below the experimental
measurements for drag resistivity in which all scattering process
are included.\cite{zheng_93,flensberg_95,hu_98} Since our calculations
already provide a very good agreement with
Kellogg {\it et al}.\cite{kellogg_02} we can argue that phonon
and impurity effects are not very important for these samples.
The phonon contribution is identified by the peak in $\rho_D/T^2$
which occurs when the average thermal phonon wave vector becomes
comparable to $2k_F$. The Kellogg {\it et al}.\cite{kellogg_02} data
do not show such a peak. Disorder in general enhances the
drag resistivity and in particular when the electron or hole
layers are close to metal-insulator transition it plays a very
important role.\cite{pillarisetty_02} We have not systematically
studied the disorder effect here but judging from the results
of our comparison to Kellogg {\it et al}.\cite{kellogg_02} data
we surmise that it is not significant.

We also note that we model the finite width of
experimental sample by an infinite square well which modifies the
bare potentials by a form factor. The effect of correct form factor
and its parameters (barrier height, etc.) obtained by well geometry
may be crucial in the final results for drag resistivity.
We have not done a self-consistent calculation of a realistic
bilayer structure to test this. Variations up to 20\% in the
quantum-well width $L$, however, does not seem to affect the
drag resistivity at low temperatures.

\section{Summary}

We have investigated the performance of various models of
inter-layer electron-electron interactions on the temperature
dependence of drag resistivity. Such models going beyond the RPA are
necessary to account for increasing correlation effects at low
density. A major input to construct an effective inter-layer
interaction is local-field corrections. We have considered the
self-consistent field approach of Singwi {\it et
al}.\cite{singwi_68} both at zero and finite temperature and FHNC
formalism to obtain intra- and inter-layer local-field corrections.
Our calculations compared with relevant experimental results of
Kellogg {\it et al}.\cite{kellogg_02} demonstrate the importance of
including correlation effects correctly in the drag resistivity
formula. The effective interaction model developed by Vignale and
Singwi\cite{vignale_85} supplemented by local-field corrections from
FHNC approach provides very good quantitative agreement with
experiments. Furthermore, previous application\cite{yurtsever_03} of
the VS effective interaction model
with simplified local-field corrections find justification
in the present calculations. In the temperature range of
Kellogg {\it et al}.  experiments\cite{kellogg_02} where the
plasmon contribution should not be significant, static
local-field corrections account for the observed drag resistivity.

It would be of interest to develop frequency dependent local-field
corrections at a similar level of sophistication presented in this
work to investigate the dynamic correlations. They are expected to
be important for the plasmon-mediated drag occurring at high
temperatures ($T\sim T_F$) as discussed by Flensberg and
Hu\cite{flensberg-hu} and most recently by Badalyan {\it et
al}.\cite{badalyan} Especially, single- and multi-pair decay
mechanisms when properly incorporated in the dynamic correlations
may explain the observed behavior\cite{nohplas} of drag resistivity
in bilayers with unmatched densities. Another possible direction is
to study the phonon-mediated drag for low density systems which
should be effective at layer separations larger than those considered
in this work.

\begin{acknowledgments}
We are grateful to G. Vignale for illuminating discussions and
comments. We thank B.Y.-K. Hu for discussions at an earlier stage.
This work is supported by TUBITAK (106T052) and TUBA.
\end{acknowledgments}

\newpage

\begin{figure}
\begin{center}
\includegraphics[width=0.60\linewidth]{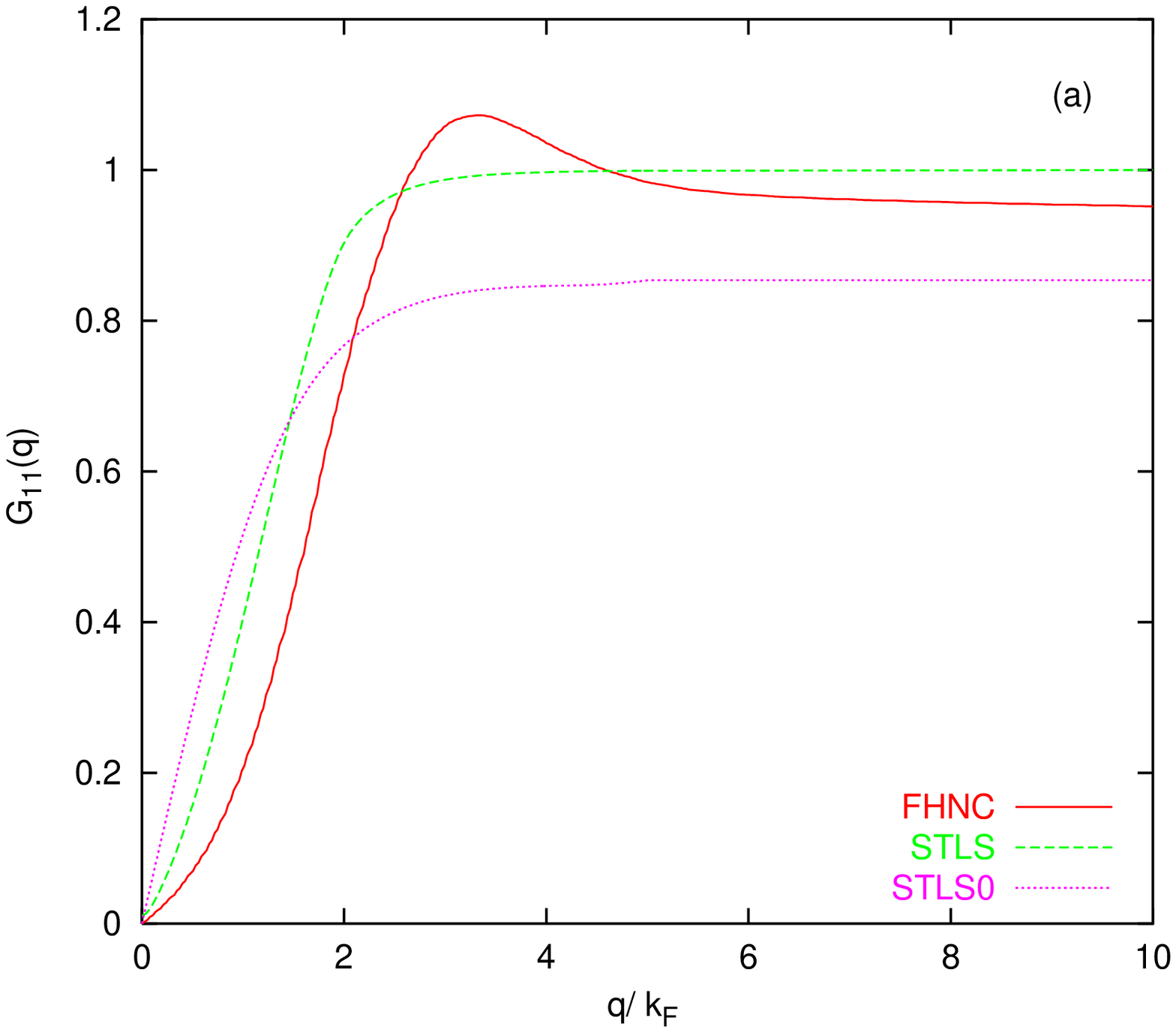}
\includegraphics[width=0.60\linewidth]{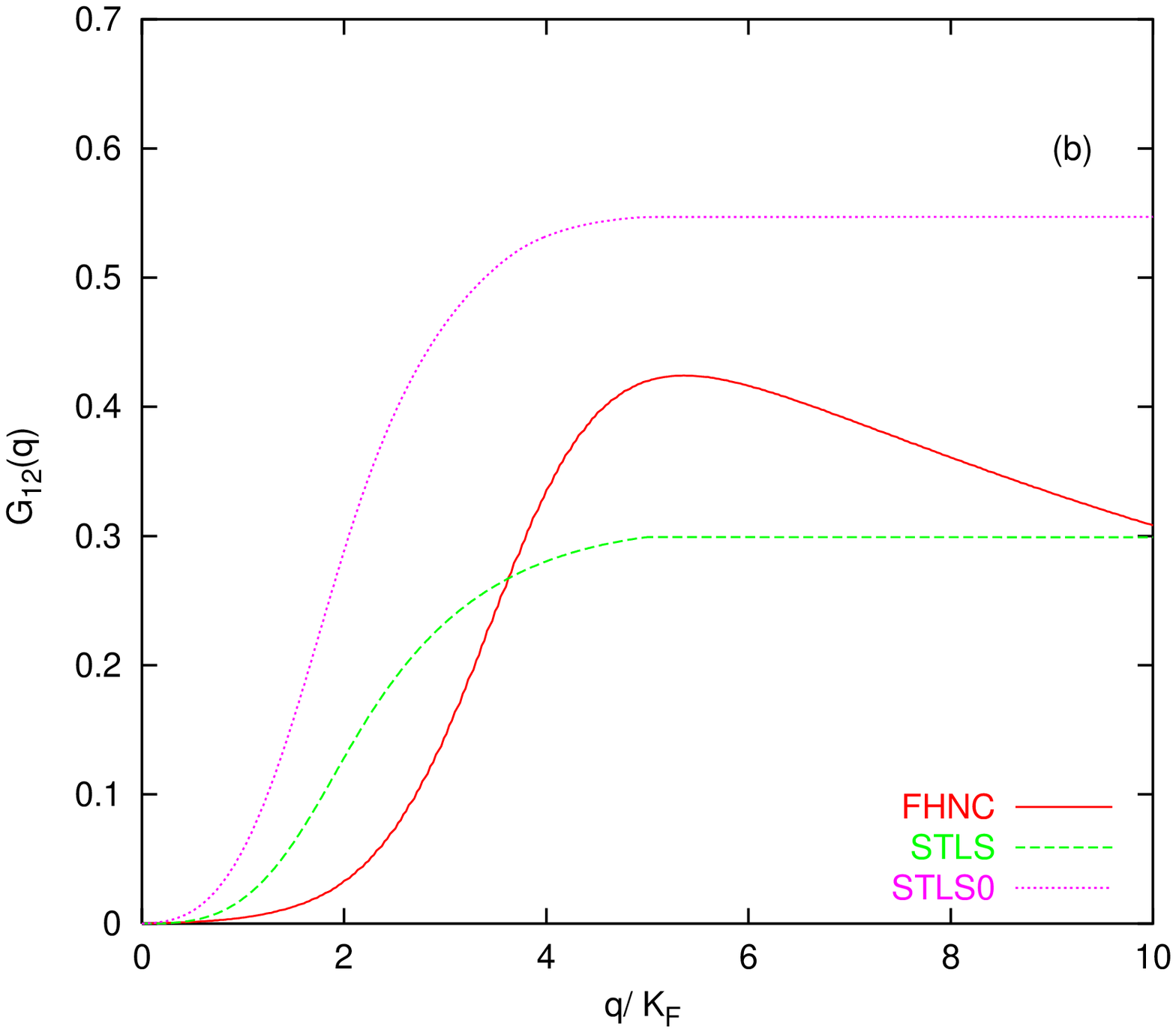}
\caption{(Color online) The local-field corrections (LFC) in various
models. (a) Intra-layer LFC $G_{11}(q)$, (b) inter-layer LFC
$G_{12}(q)$. Solid, dashed, and dotted lines correspond to FHNC,
STLS( at $T=1$\,K), and STLS0, respectively. The calculations are for
equal density layers ($n=3.1\times 10^{10}$\,cm$^{-2}$) and sample
parameters are as in Ref.\,\onlinecite{kellogg_02}.}
\end{center}
\end{figure}
\newpage

\begin{figure}
\begin{center}
\includegraphics[width=0.60\linewidth]{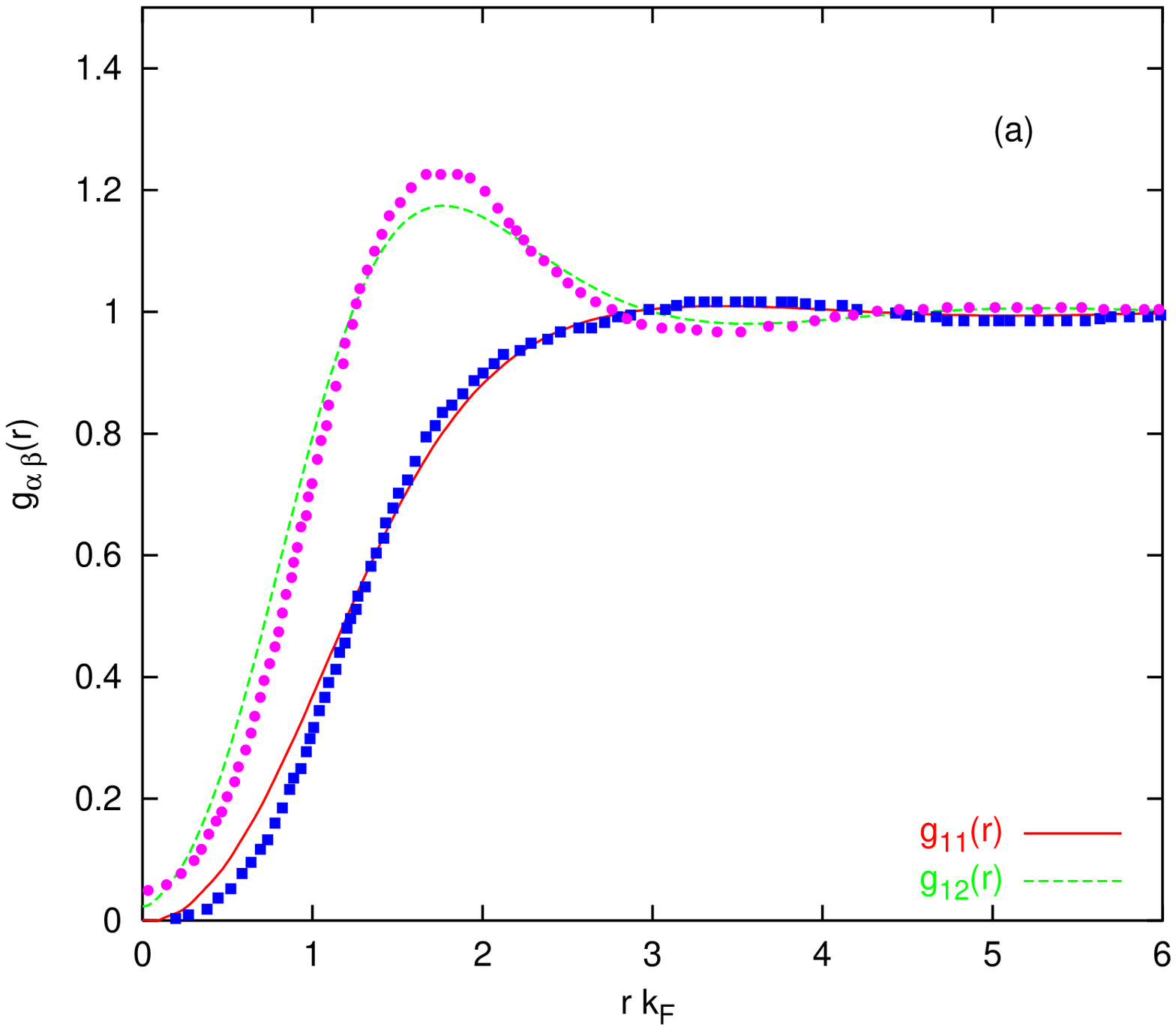}
\includegraphics[width=0.60\linewidth]{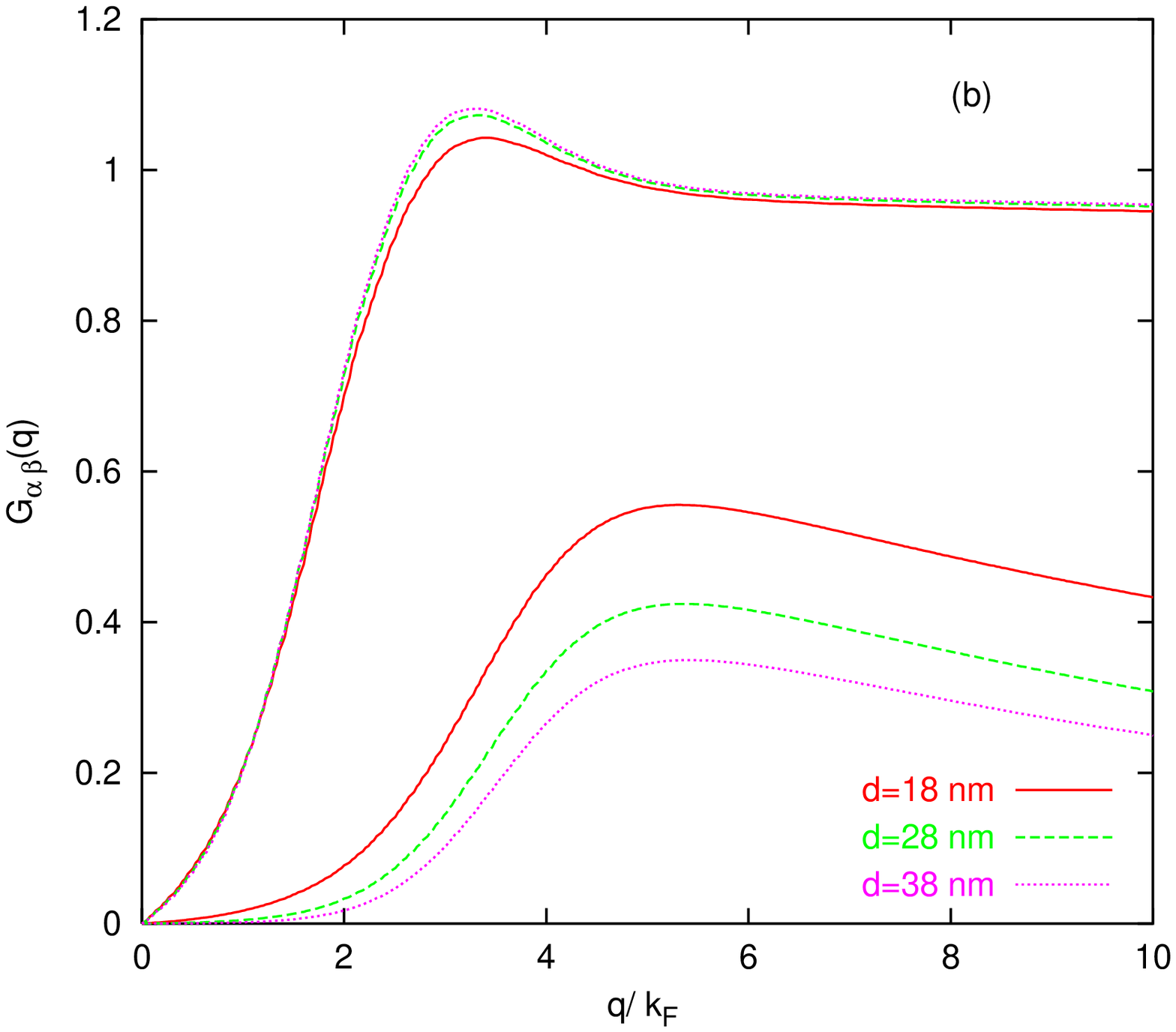}
\caption{(Color online) (a) The intra- and inter-layer pair-correlation
functions at $r_s=7.07$ calculated within the FHNC approach in
comparison with QMC results of Ref.\,\onlinecite{badalyan}
(b) The intra- and inter-layer local-field corrections (LFC) at
$n=3.1\times 10^{-10}$\,cm$^{-2}$ ($r_s=3.25$) calculated
within the FHNC approach for different inter-layer distances $d$.}
\end{center}
\end{figure}
\newpage

\begin{figure}
\begin{center}
\includegraphics[width=0.50\linewidth]{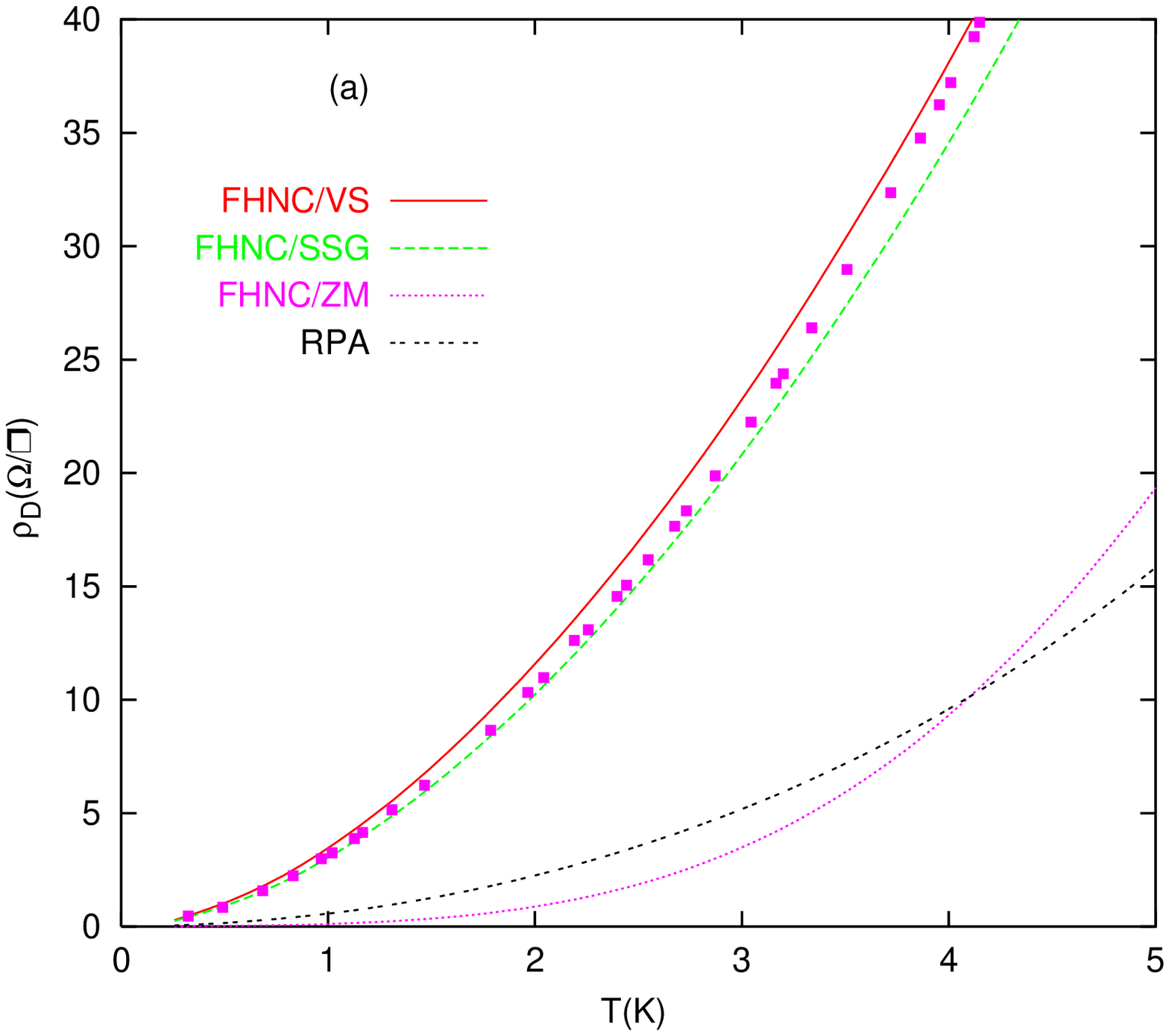}
\includegraphics[width=0.50\linewidth]{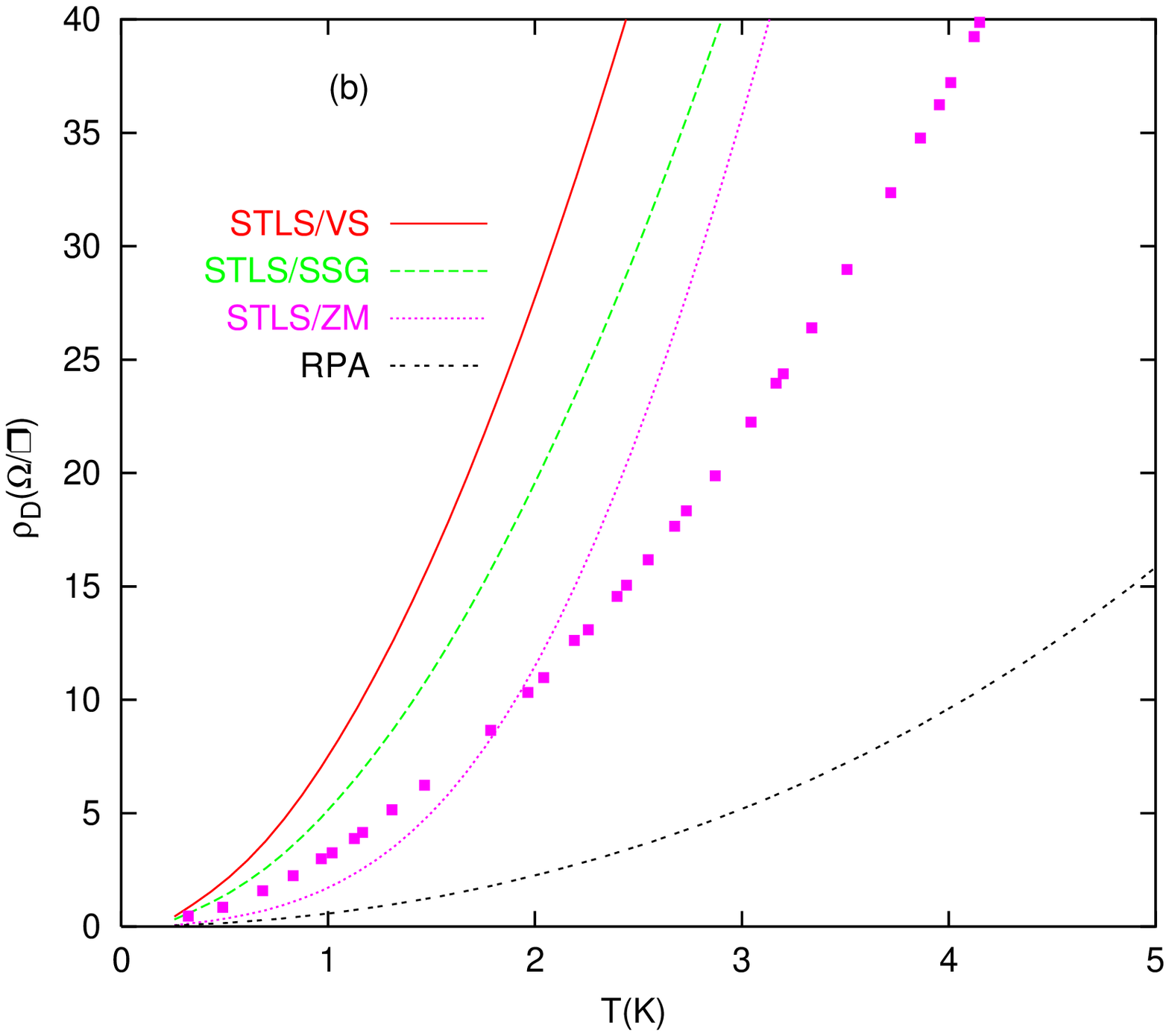}
\includegraphics[width=0.50\linewidth]{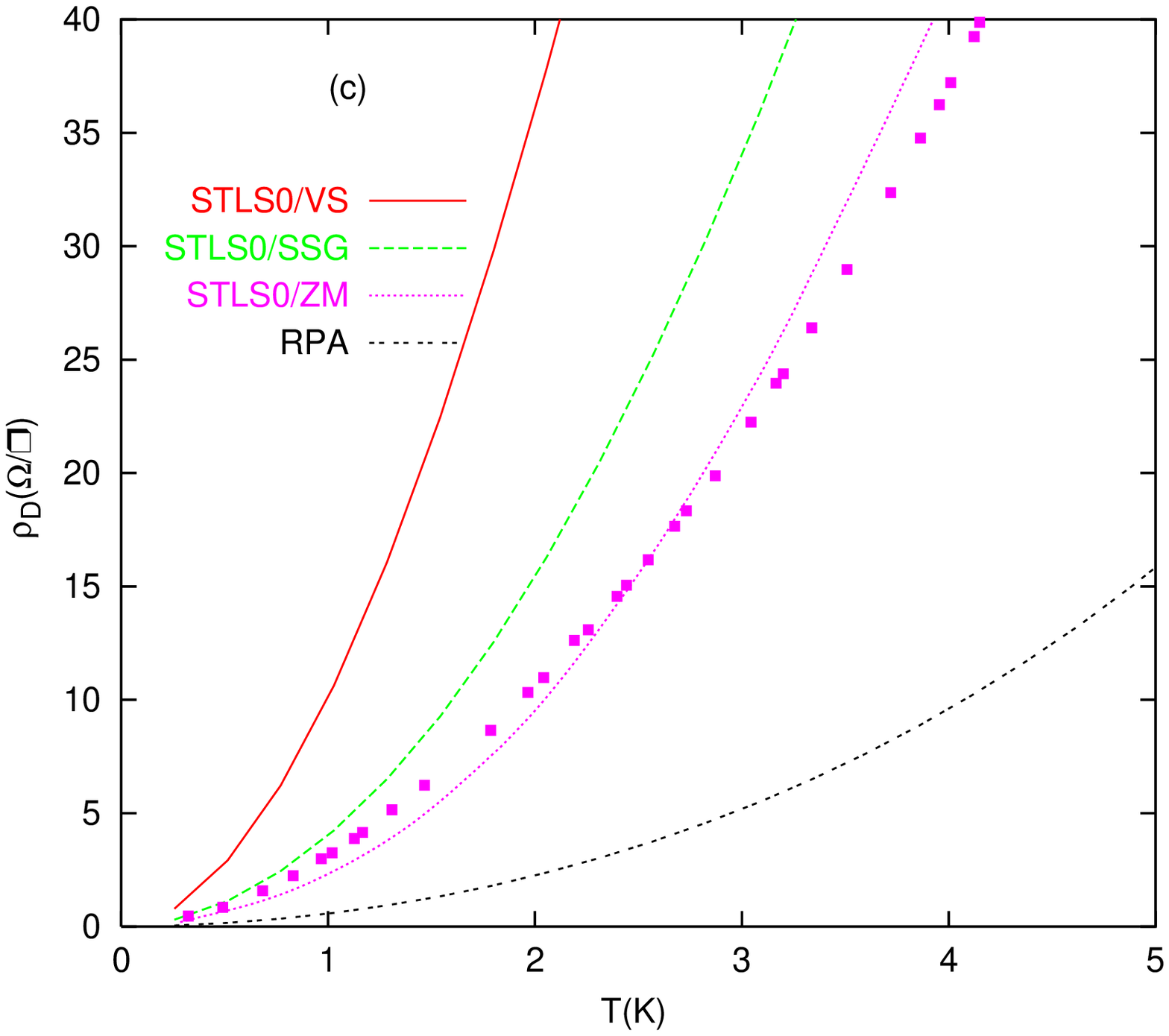}
\caption{(Color online) The temperature dependence of the drag
resistivity for the identical bilayer electron-electron systems for
$n=3.1\times 10^{10}$\,cm$^{-2}$ ($r_s=3.25$). The full boxes are the
experimental data of Ref.\,\onlinecite{kellogg_02}. (a) FHNC, (b) STLS,
and (c) STLS0 local-field corrections are used in conjunction with
different screened inter-layer interaction models.}
\end{center}
\end{figure}
\newpage

\begin{figure}
\begin{center}
\includegraphics[width=0.50\linewidth]{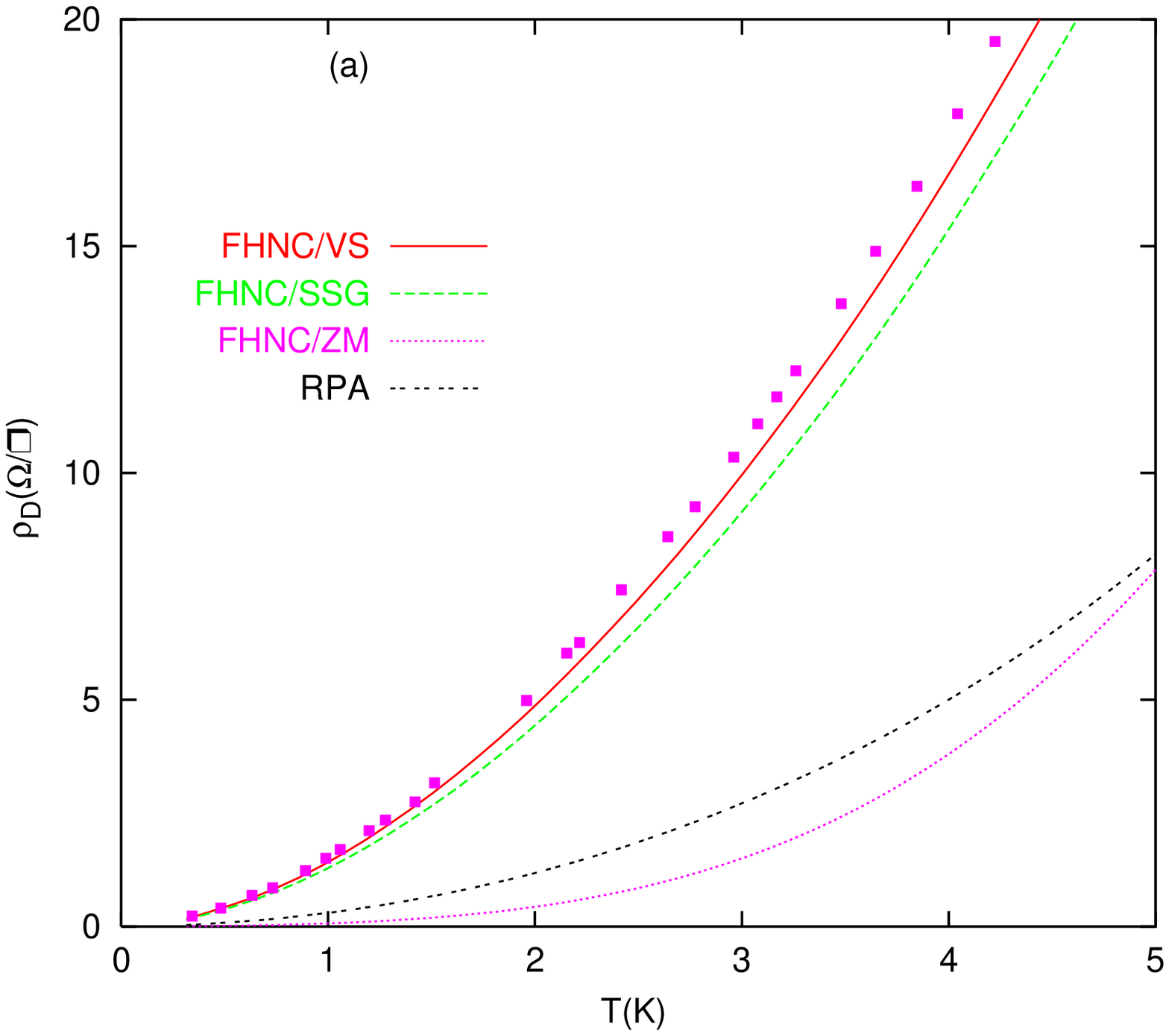}
\includegraphics[width=0.50\linewidth]{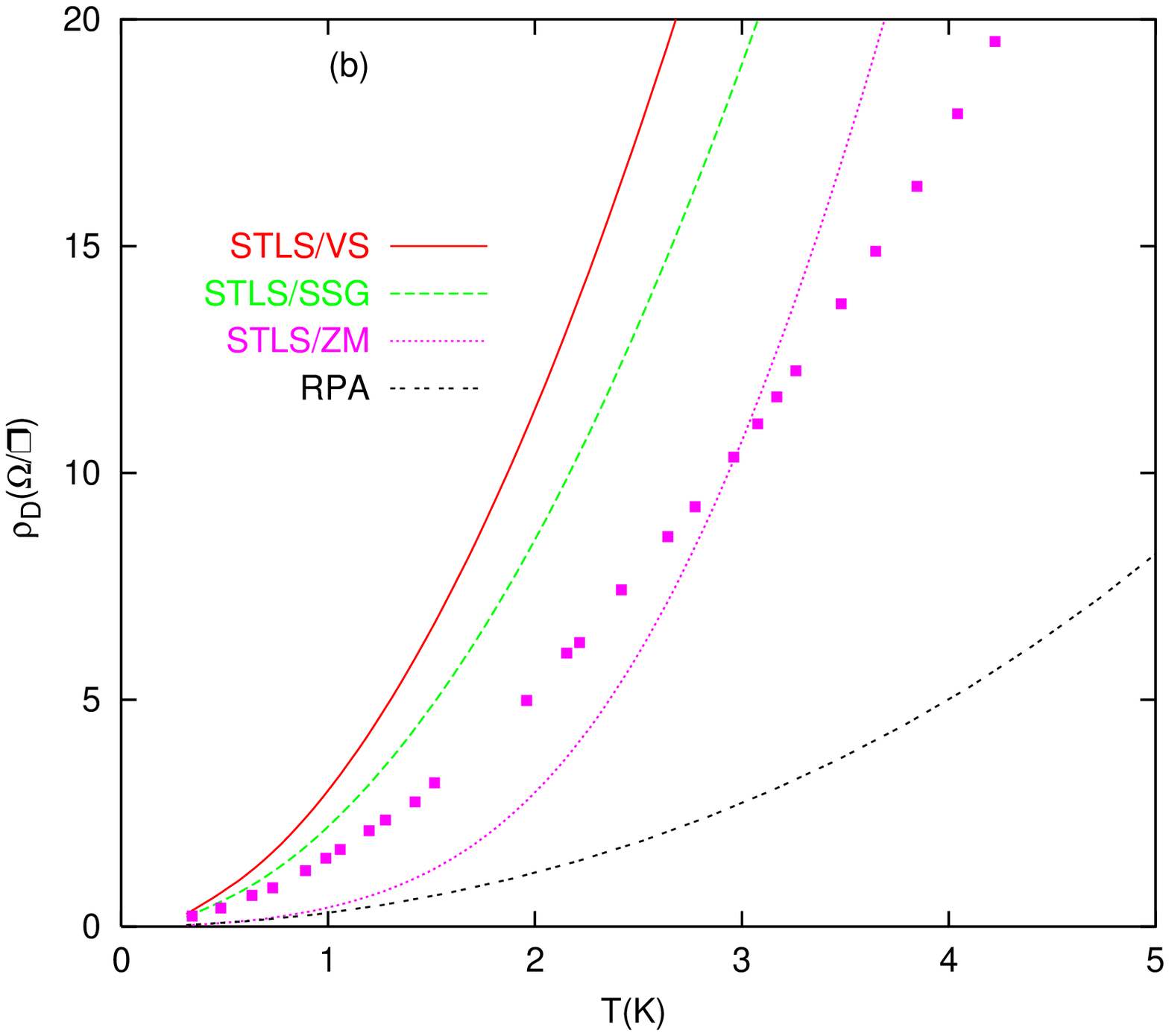}
\includegraphics[width=0.50\linewidth]{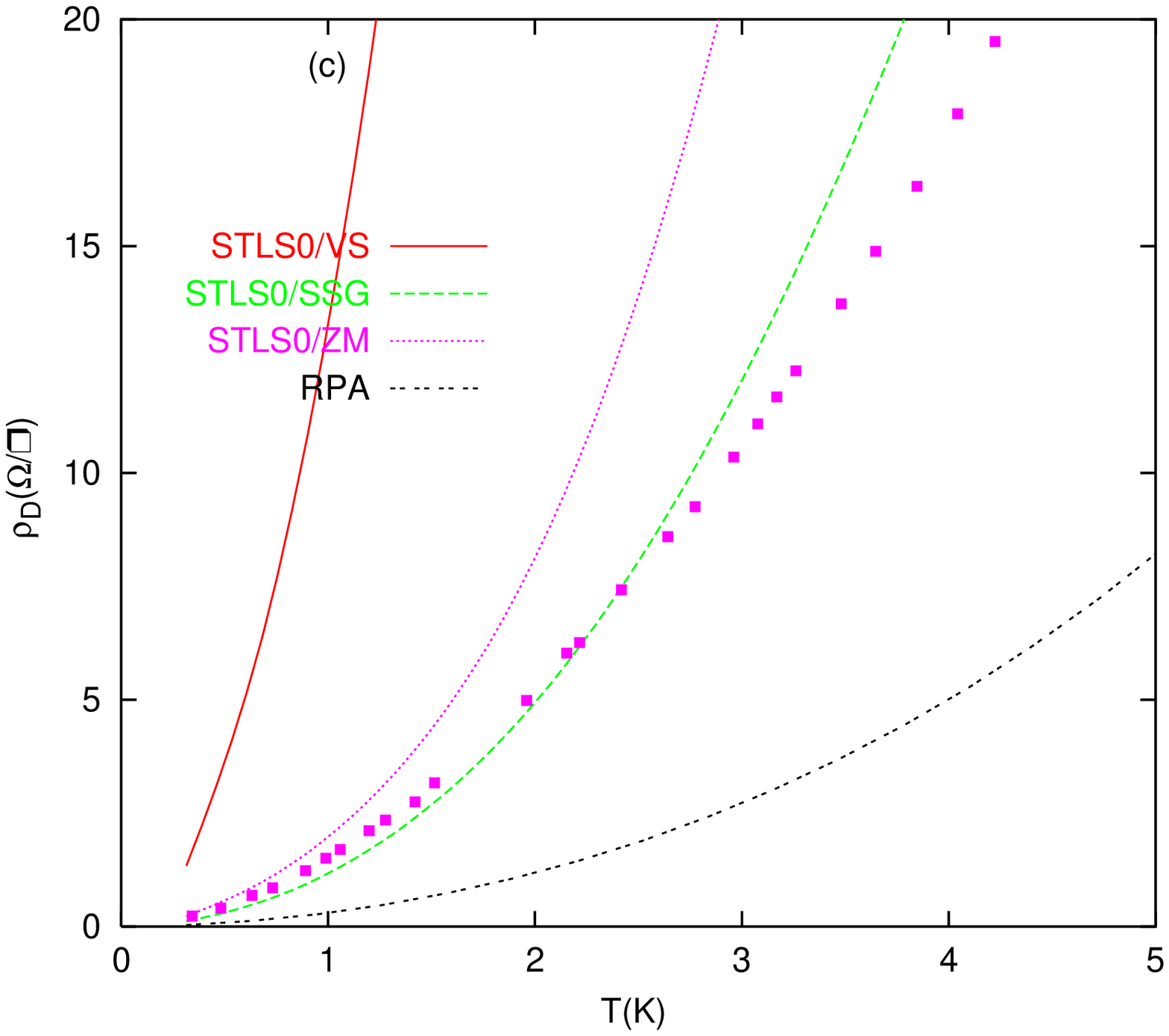}
\caption{(Color online) The temperature dependence of the drag
resistivity for the identical bilayer electron-electron systems for
$n=3.8\times 10^{10}$\,cm$^{-2}$ ($r_s=2.93$). The full boxes are the
experimental data of Ref.\,\onlinecite{kellogg_02}. (a) FHNC, (b) STLS,
and (c) STLS0 local-field corrections are used in conjunction with
different screened inter-layer interaction models.}
\end{center}
\end{figure}
\newpage

\begin{figure}
\begin{center}
\includegraphics[width=0.50\linewidth]{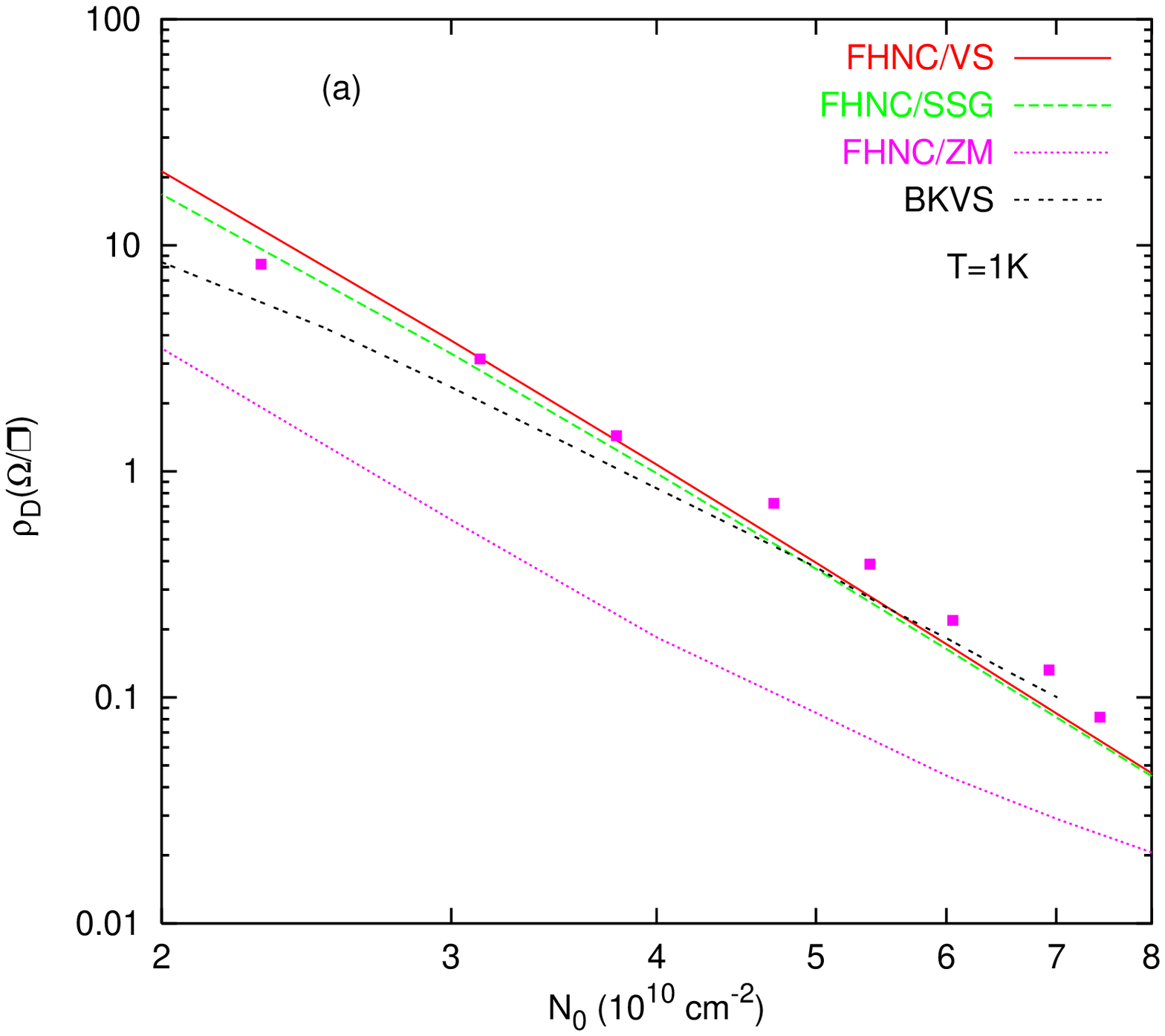}
\includegraphics[width=0.50\linewidth]{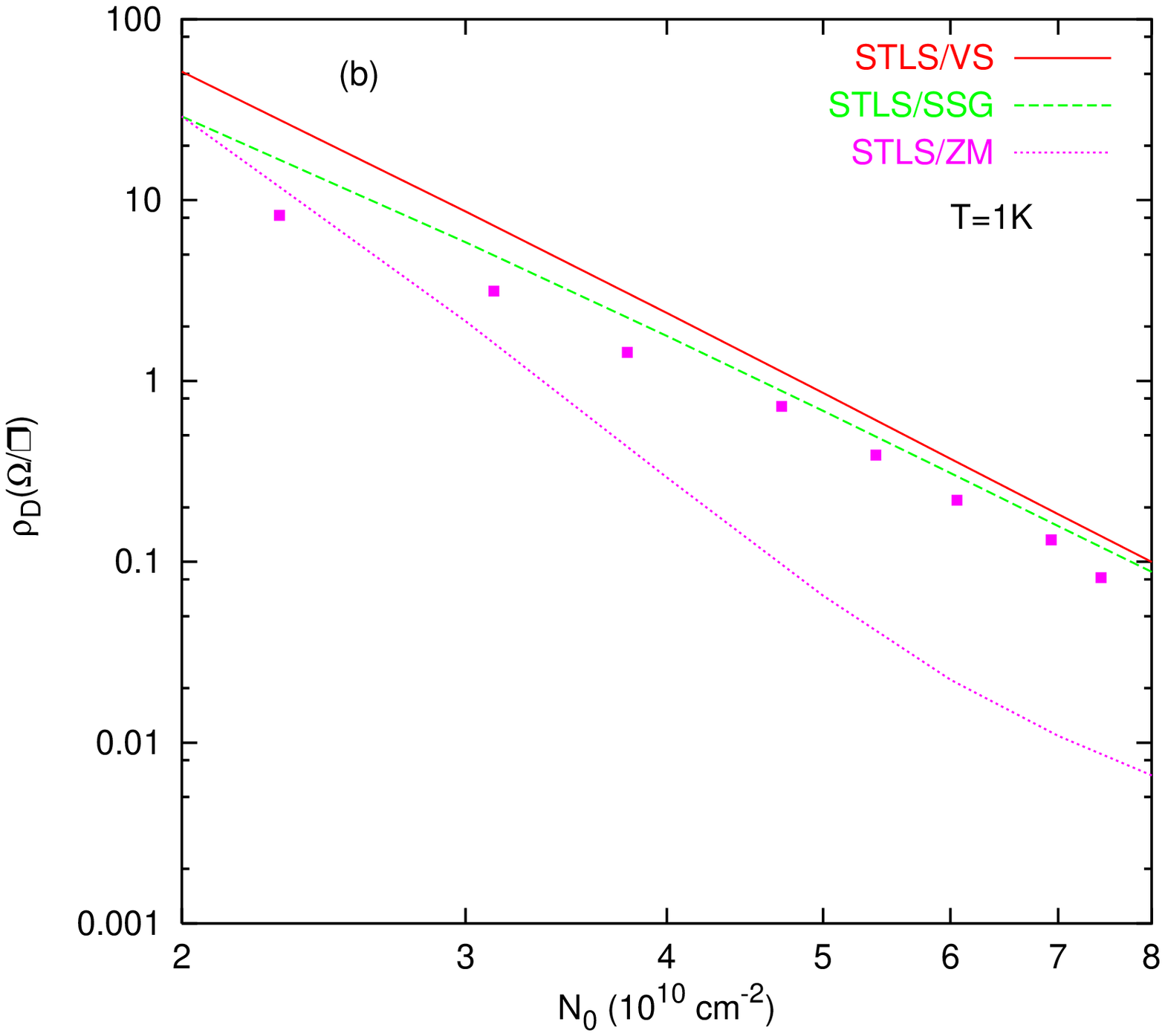}
\includegraphics[width=0.50\linewidth]{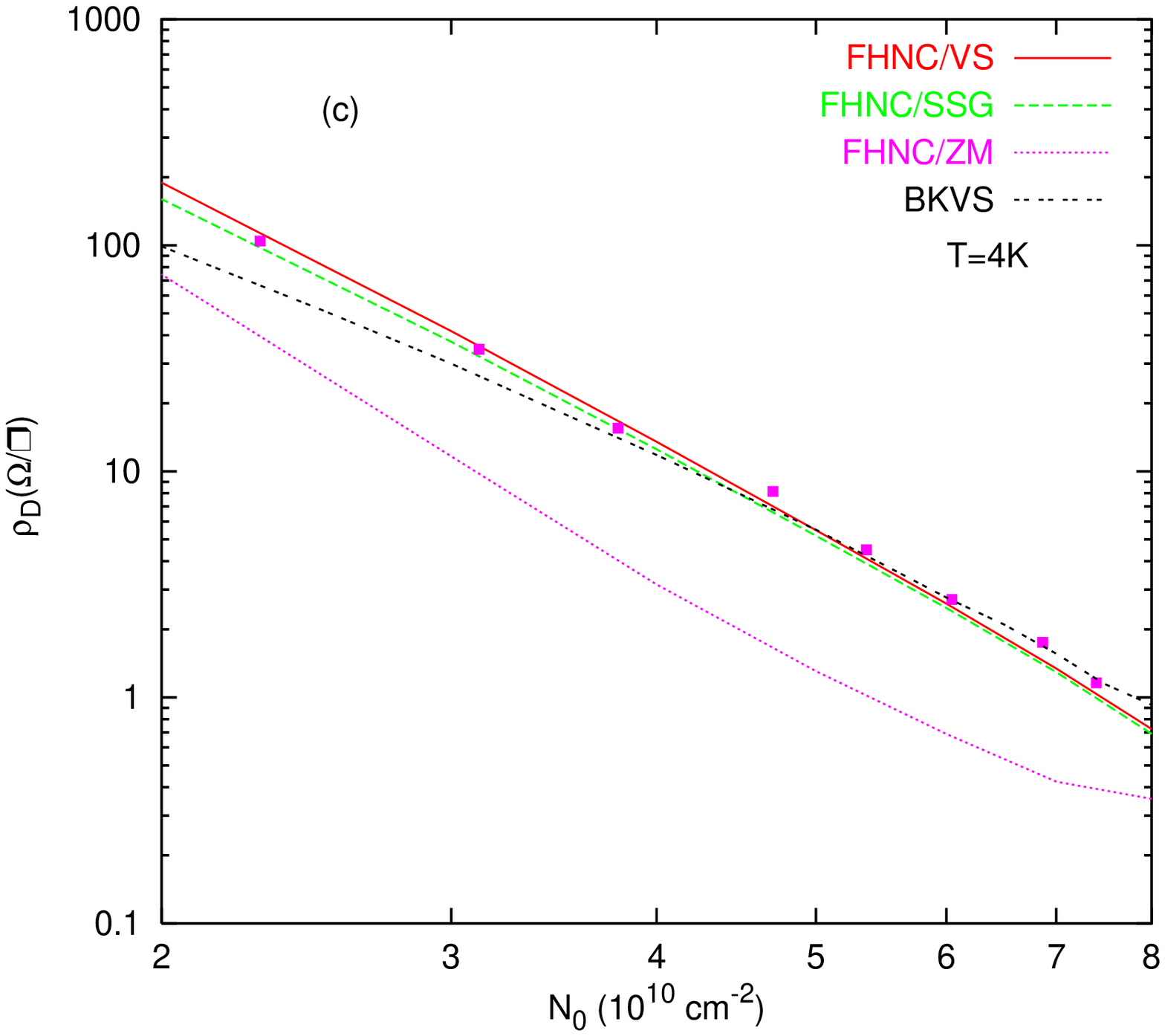}
\caption{(Color online) The density dependence of the drag
resistivity for the identical bilayer electron-electron systems at
$T=1$ and 4\,K in a log-log plot. The full boxes are the experimental
data of Ref.\,\onlinecite{kellogg_02}. BKVS refers to
Ref.\,\onlinecite{badalyan}.
(a) FHNC, (b) STLS
local-field corrections are used in conjunction with
different screened inter-layer interaction models at $T=1$\,K.
(c) Same as (a) at $T=4$\,K.}
\end{center}
\end{figure}

\newpage

\begin{figure}
\begin{center}
\tabcolsep=0 cm
\includegraphics[width=0.60\linewidth]{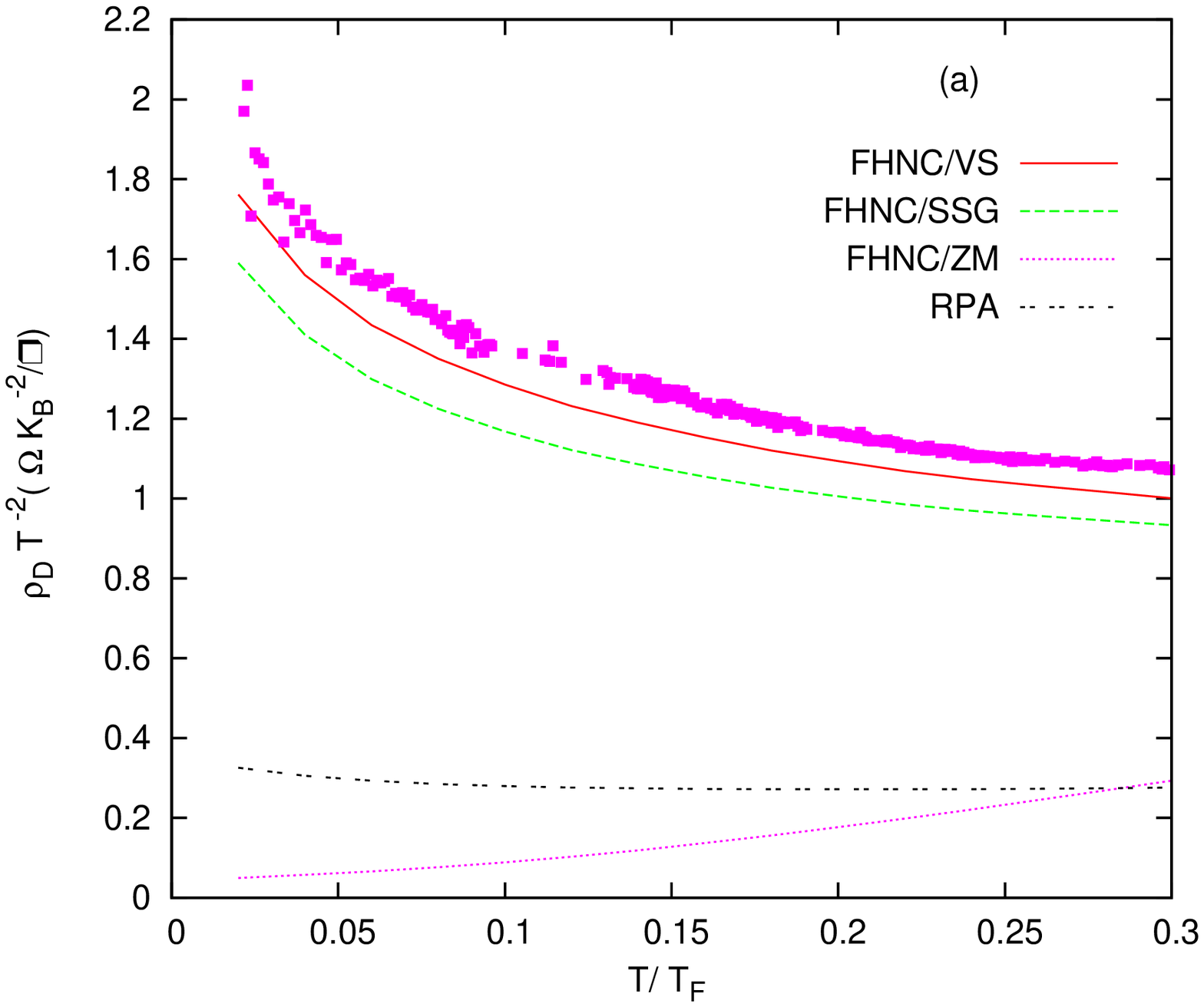}
\includegraphics[width=0.60\linewidth]{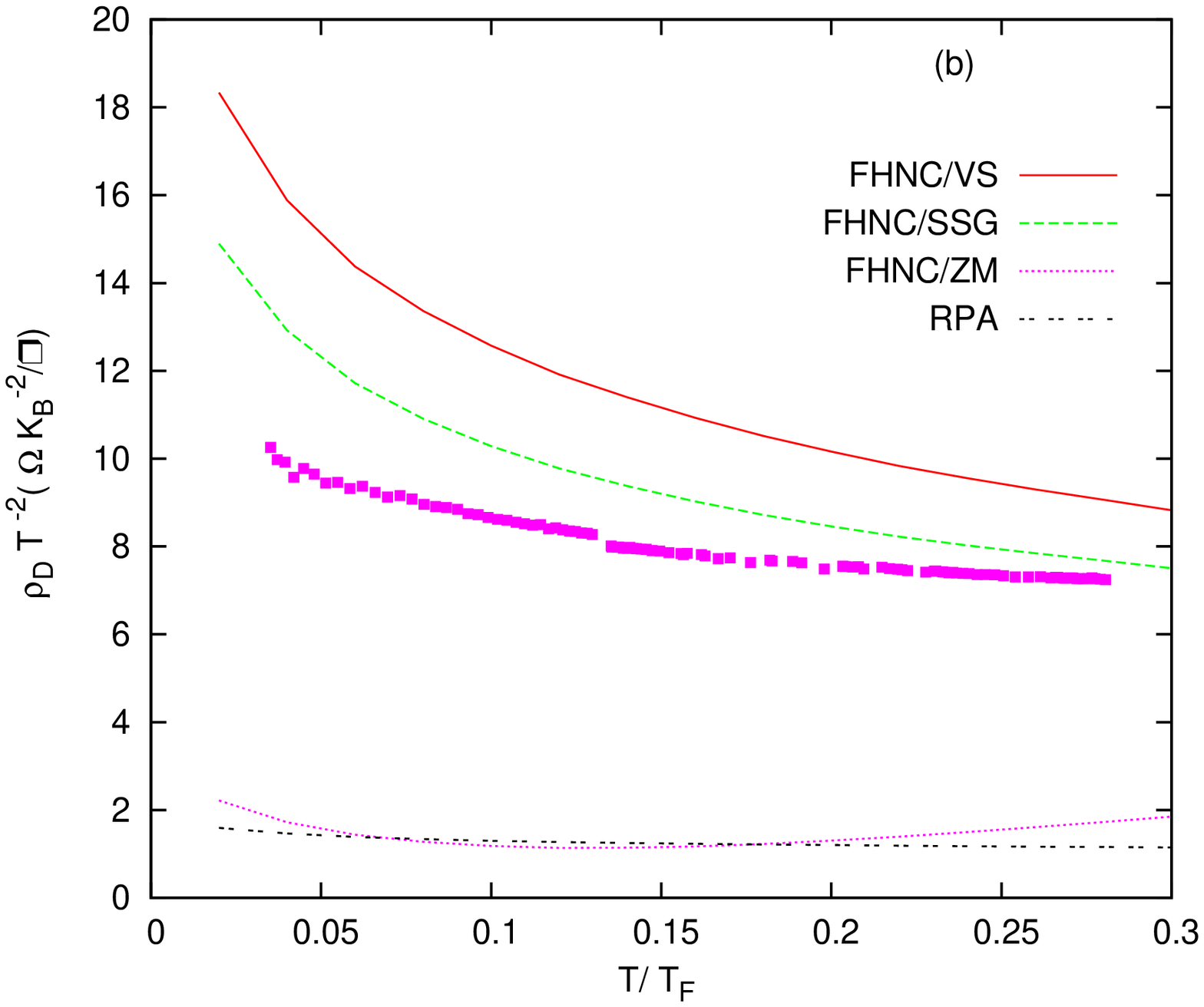}
\caption{(Color online) The scaled drag rate $\rho_D/T^2$ as a function
of temperature for (a) $n=3.8\times 10^{10}$\,cm$^{-2}$ ($r_s=2.93$) and
(b) $n=2.3\times 10^{10}$\,cm$^{-2}$ ($r_s=3.77$). The full boxes are the experimental
data of Ref.\,\onlinecite{kellogg_02}. FHNC local-field
corrections used in conjunction with different inter-layer
interaction models are compared with RPA.}
\end{center}
\end{figure}

\newpage
\begin{figure}
\begin{center}
\includegraphics[width=0.50\linewidth]{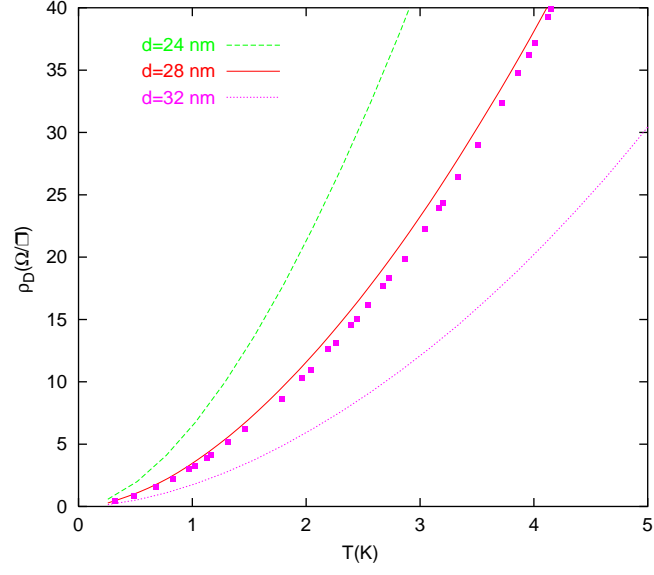}
\caption{(Color online) The temperature dependence of the drag
resistivity for a bilayer electron system with layer
density $n=3.1\times 10^{10}$\,cm$^{-2}$ ($r_s=3.25$)
The full boxes are the
experimental data of Ref.\,\onlinecite{kellogg_02} at the same
density and $d=280$\,\AA.}
\end{center}
\end{figure}

\end{document}